\documentclass[12pt]{article}

\usepackage{amsmath}
\usepackage{amssymb}
\usepackage{amsfonts}
\usepackage{amsthm}
\usepackage{setspace}
\usepackage{geometry}
\usepackage{microtype}
\usepackage{tikz}
\usetikzlibrary{trees}
\usepackage{istgame}
\usepackage{graphicx}
\usepackage{caption}
\usepackage{colortbl}
\usepackage{threeparttable} 
\usepackage{dsfont}
\usepackage{multirow}
\usepackage{float}
\usepackage{bbm}

\theoremstyle{definition}

\title{Hiring Intrinsically Motivated Agents: A Principal's Dilemma}
\author{Andrew Leal}

\begin{document}

\maketitle
\begin{abstract}
    Employers are concerned not only with a prospective worker's ability, but also their propensity to avoid shirking. This paper proposes a new experimental framework to study how Principals trade-off measures of ability and prosocial behavior when ranking Agents for independent jobs. Subjects participate in a simulated, incentivized job market. In an initial session, subjects are Workers and generate a database of signals and job results. Managers in subsequent sessions observe the signals of Worker behavior and ability and job details before a rank-and-value task, ranking and reporting a value for each Worker for two distinct jobs. Results highlight Managers' preference for ability over prosocial behavior on average, especially for Managers in STEM fields. There is evidence of homophily: the relative value of prosocial behavior is higher for highly prosocial Managers, compensating for ability or even surpassing it in value. 
\end{abstract}

\newpage

\begin{quote}
    It was the face of a man of hardened integrity who could no more consciously violate the moral principles on which his virtue rested than he could transform himself into a despicable toad.
\end{quote}
\hfill --Joseph Heller, Catch-22

\section{Introduction}

Shirking is a concern for any employer. Unwanted behavior could be mitigated using financial incentives, but constraints often render contracts incomplete in practice. For example, an existing collective bargaining agreement may limit the financial incentives a firm can offer, and the recent shift towards remote work reduces employee visibility. Furthermore, using external incentives or controls to influence effort may undermine trust (Falk and Kosfeld, 2006; Weber and Mayer, 2011; Lumineau, 2017) or crowd-out psychological rewards from the effort (Frey, 1997; Kreps, 1997; B\'enabou and Tirole, 2003, 2006), resulting in worse outcomes relative to the absence of these incentives. Given the complex interaction between external incentives and effort, a more straightforward approach to reducing shirking would be to prioritize individuals intrinsically motivated to avoid shirking. To the extent that internal motivations for this prosocial behavior are present, individuals will be steered away from opportunistic behavior. Yet, in light of these considerations, productive ability remains a vital predictor of job performance. This paper empirically investigates the trade-off employers face between prioritizing candidates' past behavior and productive ability, specifically focusing on prosocial behavior versus talent when evaluating candidates for an independent task. 

Organizations increasingly recognize that employees contribute value not only through productive ability but also through prosocial activities. Levashina et al. (2014) note the rise in personality-oriented hiring over time, citing the positive relationship between personality and job performance and the relatively small adverse impact of this selection style. Conscientiousness, characterized by reliability and self-discipline, was the most commonly assessed personality trait. Neglecting personality traits may bear significant costs. Replacing toxic employees — those who engage in activities that are \textit{harmful} to a firm's interests — can be more than twice as costly as the benefits from hiring a ``superstar'' (Housman and Minor, 2015). Employees intrinsically motivated to act as good stewards are less likely to engage in counterproductive behaviors, thereby providing substantial value to employers. While firms would clearly prefer prosocial superstars, the hiring manager faces a trade-off between ability and behavior when evaluating candidates with differing strengths. How is the value of desirable behavior weighed against productive ability? How do the evaluator's own behavior and ability shape their valuations of these traits in potential hires?

This paper introduces a novel experimental design to study behavior-ability trade-offs and presents evidence on the trade-off between prosocial behavior and talent when evaluating workers, exploring whether past behavior can compensate for, or even surpass, traditional measures of competence in evaluators’ assessments. Subjects participating as Managers observe measures of prosocial behavior and ability for a set of Workers, rank these Workers for simulated jobs, and then report an incentivized valuation of each Worker for the jobs before their earnings are realized. Prosocial behavior is measured as the number of tokens shared in a Dictator Game, while ability is measured as the number of correct solutions in an addition task. Jobs combine these two measurement tasks, having Workers complete addition problems for Managers, but with a choice of how many to solve. These tasks are done independently and there is no teamwork; Managers simply assess each Worker's potential, taking into account the number of problems the Worker will answer and how successful they will be in answering them. Reported values represent the minimum number of points - the experimental currency - the Manager would accept in lieu of having their earnings determined by the given Worker's performance in the jobs.

The experiment features two distinct jobs performed by subjects participating as Workers in an initial session. Each job pays a piece rate to the Worker for their output, which is uncertain and produces earnings for Managers, and differs only in the available returns to shirking. In the No Conflict job (NC), shirking is not rewarding, so financial incentives are sufficient and entirely crowd-out any intrinsic motivation against shirking. Thus, prosocial behavior should have no impact on Worker valuations. In the Conflict job (C), shirking produces earnings for the Worker, but not the Manager. To the extent that psychological benefits from intrinsically motivated actions substitute financial incentives, prosocial Workers will be less likely to shirk in C. Hence, the hiring dilemma is present in C, and Managers must consider the interplay between prosocial behavior and ability when ranking Workers. 

Relative to simply generating fictional workers, using real subjects mitigates the concern that individuals will perceive the game as mathematical in nature. Real subjects bring variability and unpredictability, making the Manager's task more comparable to real evaluation contexts where workers display unique, individual qualities. Using real workers also enables the study of not only how Managers weigh prosocial behavior and ability, but also how they interpret or attribute meaning to these signals. For example, Managers may form judgments about a Worker’s personality based on prosocial actions or make assumptions about their competency based on actual performance. These mental models incorporate emotional and social perceptions, which often play significant roles in how people assess others’ qualities. This interpretive layer is crucial for understanding how people make complex evaluations, which would be missing if the signals were artificial.

The effect of prosocial behavior and ability on reported values is analyzed using nonparametric and individual-fixed-effects regressions. On average, subjects in the experiment valued ability substantially more than past prosocial behavior. As expected, in NC, valuations are driven almost entirely by Worker ability. A small effect from prosocial behavior remains despite sufficient incentives, but is attenuated when controlling for Managers' characteristics. In C, ability was still preferred, especially by STEM participants. In the richer specifications, the marginal value associated with one more correct response in the addition task was approximately 3 times that of sending one more token in the Dictator Game, increasing to just over 8 times more for participants in STEM fields. Males valued the prosocial component more than others, but still less than they value ability. Heatmaps for job C show that an extra unit in ability is more valuable than an extra unit in the prosocial measure at almost all characteristic pairs in the space. These findings differ from Raihani and Barclay (2016) and Eisenbruch and Roney (2017), who find generosity to have a stronger impact on partner choice when contingent rewards for effort are absent. Explicit contracts, which are absent in their studies, provide a degree of assurance against shirking, reducing the value of good behavior and tipping the scale in favor of talent. In Eisenbruch and Roney (2017), productivity tended to matter more when it was associated with skill rather than luck, and it seems plausible here that skill is the most salient way a Worker can produce benefits.   

Distributions of Worker valuations highlight a substantial disagreement on the values of Workers with disparate characteristics, especially for low-prosocial, high-ability Workers. Several subjects ranked Workers with top ability scores and bottom prosocial scores in their top 3, while these same Workers appeared at low ranks for many other subjects. A similar but less pronounced pattern is observed for high-prosocial, low-ability Workers. There is rank agreement for Workers with high (low) levels in both measures, with Managers consistently ranking them among the top (bottom). These patterns are consistent with the social perception literature, where individuals perceived as warm and competent are uniformly regarded positively, those lacking warmth and competence uniformly regarded negatively, and individuals high in one dimension but low in the other eliciting mixed feelings (Fiske et al., 2007). 

What could drive such disagreement? When assessing Workers, there may be a natural bias for individuals to favour Workers with qualities similar to themselves (Mcpherson et al., 2001). There is evidence for a likeness bias on the behavioral dimension; Managers with high prosocial measures place relatively more value on this component relative to low prosocial Managers. When returns to shirking are present, Managers with prosocial measures in the top-third value the behavioral component 76\% more than Managers in the bottom two-thirds. Those with higher prosocial scores than ability scores valued the behavioral component by approximately 126\% more than Managers whose ability score was at least as high as their social score. This pattern holds even in NC, suggesting any value placed on prosocial behavior in this job is driven by these prosocial Managers. I find no evidence of this likeness bias among ability, as it appears all Managers value able Workers regardless of their own ability. 

\subsection{Literature Review}

Actions may be intrinsically motivated by an enjoyment for the work itself, the pleasure of working with particular people, or a responsibility for one's actions driven by duty, guilt-aversion, fairness considerations, or other moral concerns (see Battigalli and Dufwenberg (2022) for a review on psychological games). This paper is concerned with the last type and considers intrinsic motivation to behave prosocially, as in B\'enabou and Tirole (2006). 

This paper contributes to the literature on partner choice, revealing a preference in Worker ability over past prosocial behavior. Although psychologists have been concerned with the role of prosocial behavior in partner choice for decades, studying the value of a partner's productive ability is a relatively new area in this literature (Macfarlan and Lyle, 2015). Fiske et al. (2007) identify warmth and competence as the two universal factors in social perception. Aligned with the findings in this paper, their work highlights the uniform positive and negative perceptions of those considered either having or lacking both warmth and competence, with individuals high in one dimension but low on the other eliciting mixed feelings. Raihani and Barclay (2016) use a modified Dictator Game to study partner choice when prospective partners differ in wealth and generosity. Their experiment features two stages: in the first part, individuals are endowed with either \$0.50 (``poor'') or \$2.50 (``rich''), representing differences in ability to share surplus, and can share either 20\% (``stingy'') or 50\% (``fair'') of this surplus. In a second round, a third subject observes the endowments and share amounts by two dictators and chooses one to play receiver to. More than half of the participants favor partnership with poor-fair types than rich-stingy types, though results are not statistically significantly different from chance. Eisenbruch and Roney (2017) study productivity and generosity using Trust Game experiments. In the first stage, subjects send a portion of their \$10 endowments to an anonymous partner. The amount is multiplied by either 3, 4, or 5, representing variations in productivity, and sent to a trustee. Trustees can return either 30\%, 40\% or 50\% of the total amount received. Generosity had a larger impact on partner choice and perceived fairness than did productivity. They note that the effect of productivity was stronger when associated with skill rather than luck, though it still had a lower impact than generosity.

While these studies provide evidence for how ability and prosocial behavior may be weighed when choosing partners, the presence of a contract complicates relationships, influencing expectations and behavior in complex ways. To the extent that payments for outcomes transforms a partnership into an economic relationship, financial incentives may crowd-out an Agent's intrinsic motivation (Deci, 1971; Frey, 1997; Frey and Oberholzer-Gee, 1997; B\'enabou and Tirole, 2003, 2006). Prevention-framed contracts, which use external incentives or controls aimed at directing behavior, signal a distrust in the Worker, substituting intrinsic motivations (Weber and Mayer, 2011; Lumineau, 2017). Falk and Kosfeld (2006) show that signaling this distrust is costly; when a Principal can impose minimum effort levels on their Agents, aggregate effort decreased toward the minimum. The experiment in this study explicitly considers a Manager-Worker relationship, rather than friendships or other types of cooperative relationships, and incorporates a prevention-style contract with incentives for supplying effort. Compared to Raihani and Barclay (2016) and Eisenbruch and Roney (2017), using the language of Manager and Worker adds meaning, potentially evoking different feelings than when evaluating ``partners'' or ``players'' (Alekseev et al., 2017), and Workers needed to provide real effort to produce earnings for Managers, with incentives attached to this effort.  

The experimental design differs from existing studies on multi-dimensional partner choice in several ways. In Raihani and Barclay (2016) and Eisenbruch and Roney (2017), measures are restricted to 2 and 3 outcomes, respectively. These limitations make it difficult to uncover the rate of substitution between the two forces. This design uses measures of prosocial behavior and ability that each span up to 11 discrete outcomes, enabling the study of how increases in one characteristic impact worker valuations at many different characteristic pairs. The design also adds economic context to the relationship.  Furthermore, the trade-off between behavior and ability should be influenced by any contracts underlying the relationship, a key feature here. In NC, financial incentives are sufficient to deter Workers from shirking, so that a payoff-maximizing Manager should value only Workers' ability. The dilemma is present in C, where even the highest ability Workers can earn more from shirking.

This paper also explores the role of one's own traits when assessing employee characteristics, adding to the literature on homophily. Homophily is an important mechanism underlying partner choice, and has been documented across a broad spectrum of relationships and over several characteristics such as age and race (McPherson et al., 2001). In the economics literature, evidence of homophily exists when referring and hiring job candidates (Burks et al., 2015) and in social networks (Currarini et al., 2009; Currarini et al., 2016). This pattern tends to be explained by biases, both in preferences and in meetings (Currarini et al., 2009). Following this literature, the analysis in this paper investigates whether subjects place relatively more value on Workers with similar characteristics. Such behavior would be driven by biases in preferences, since the experimental design eliminates any biases in meeting opportunities that may arise from endogenous job application decisions. 

This paper is organized as follows. The next section describes the experimental environment, design, and implementation in detail. Results from the experiment are found in Section 4 before a discussion of design choices and future directions in the final sections. 

\section{Experiment}

\subsection{Conceptual Framework}

A Principal evaluates Agents from a candidate pool. Each Agent $n$ has characteristics $(x_n,y_n)$. The trade-off in characteristics when evaluating job candidates can be viewed as a trade-off between ``goods'', where each characteristic increases the expected utility from hiring a particular Agent and characteristic bundles $(x_n,y_n)$ are analogous to consumption bundles. The Principal has a preference relation $\succ$ where $n \succ n^{\prime}$ implies that qualities $(x_n,y_n)$ are preferred to $(x_{n^\prime}, y_{n^\prime})$. Agents produce output $\pi$, which is uncertain and depends on the Agent's characteristics. Principals form expectations of the utility a given Agent provides. Each expected utility level is associated with a certainty equivalent $V_n$ such that
\begin{align*}
    E\big[u(\pi) \lvert x_n, y_n\big] &= u\big(V_n\big) \\
    \implies \frac{\partial E[u(\pi) \lvert x_n, y_n]}{\partial z} &= \frac{\partial u}{\partial V_n} \frac{\partial V_n}{\partial z} \ \text{where } z \in \{x_n,y_n\} \\
\end{align*}

Thus, while $u(\cdot)$ is never observed, values $V_n$ provide cardinal information on the rankings of Agents. That is, if $V_n - V_{n^\prime} = c$ for some constant $c \in \mathbb{R}^{+}$, then $c$ represents the extra money a Principal would need to hire Agent $n^\prime$ over Agent $n$, with utility difference $u(c)$. In other words, this Manager is indifferent between hiring Agent $n$ and hiring Agent $n^\prime$ when they also receive an additional $c$ units of money for hiring $n^\prime$.

\subsection{Overview}

Subjects participate in a simulated job market, playing the role of either a Worker (Agent) or a Manager (Principal). The experiment consists of one Worker session and multiple Manager sessions, each containing three parts. The Worker session is conducted first, creating a database of Workers to be evaluated by subjects in the subsequent Manager sessions. Prosocial behavior and ability are measured in the first two parts of each session. Prosocial behavior is measured by an allocation decision in a modified Dictator Game, and ability is measured by performance in a simple addition task. In the third part of the Worker session, Workers perform two jobs. In the third part of the Manager sessions, Managers evaluate the Workers for these jobs, given the Worker's allocation decision in the Dictator Game and the number of correct responses in the addition task. Subjects accumulate points in all three parts, which are converted to Canadian dollars at a rate of \$0.08 per point and paid in cash at the end of the session.

The goal of this design is to create an environment where Workers are evaluated on two important but distinct dimensions, controlling for other differences across Workers, such as recruitment costs,  contracts, and the signal-generating process. Signals are provided at no cost and are easy for Managers to interpret; each is an integer between 0 and 10, generated by tasks that Managers also independently complete. Each Worker's signal pair comes from the same tasks, each Worker performed the same jobs, and each Manager evaluated the same set of Workers for these jobs. This controls for unobserved differences in beliefs about how signals are generated and elicits many valuations for each Worker in the same job environment. To alleviate the concern that Managers view the signals as being strategically generated, subjects received instructions for each part as they were played so they could not condition choices on future parts. The order in which the measurement tasks are presented is chosen to further reduce strategic behavior across tasks. Instructions for the Worker session state that the decisions made will be used in future sessions, but did not explicitly state that Workers would be evaluated on their signals. In addition, Worker payoffs do not depend on the Managers' evaluations, but only on the job outcomes. Hence, these endogenous signals are not generated strategically, but rather by innate traits. This degree of control focuses the task of evaluating Workers on their signals, and thus elicits preference orderings over the two characteristics under varying job parameters. Other advantages of a laboratory experiment in this context is that it grants complete control over the job parameters, inducing easily interpreted bounds on payoffs and allowing for a simple worker history to be produced.

\begin{figure}[h]
    \centering
    \includegraphics[width=.5\linewidth]{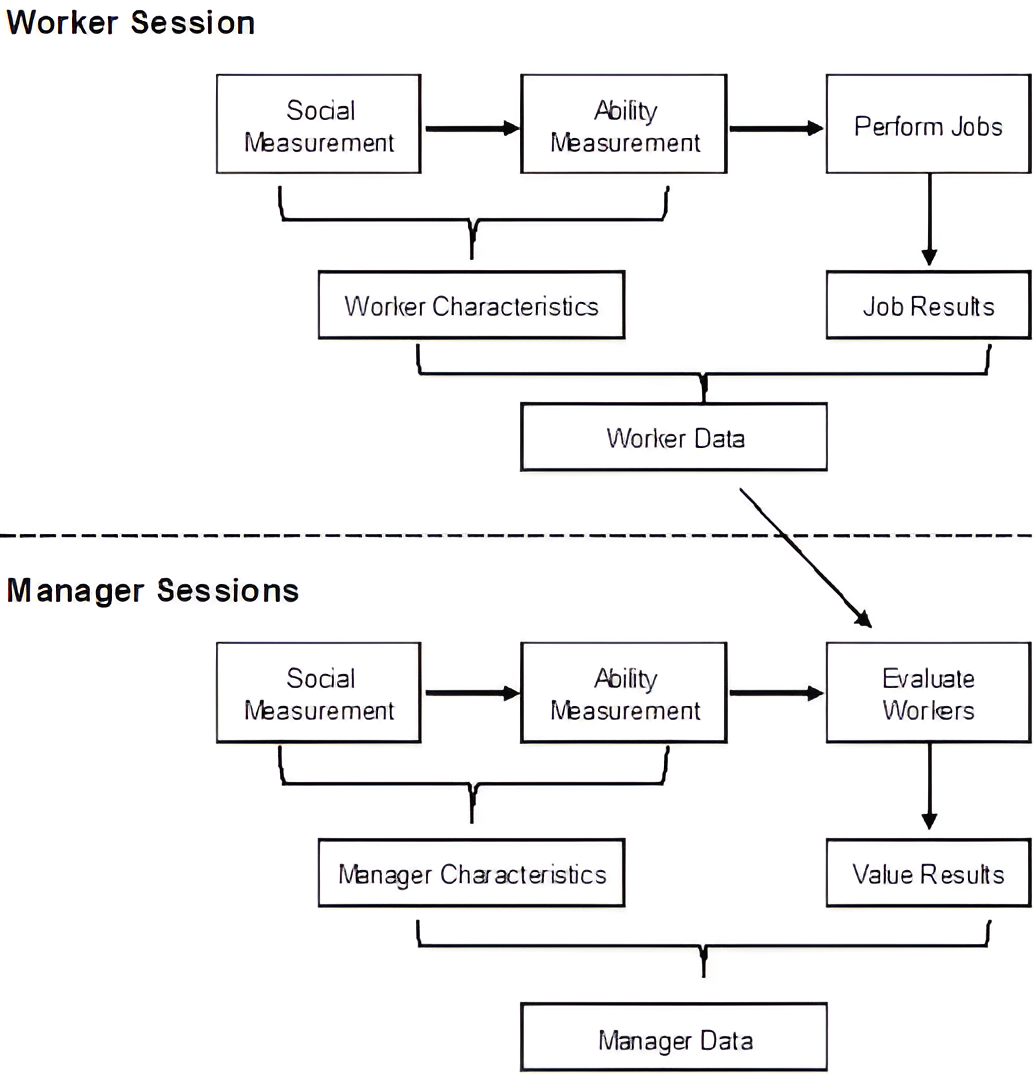}
    \caption{Flowchart of Experimental Sessions}
    \label{fig:flow}
\end{figure}

The Dictator Game and job task include incentivized quizzes prior to starting. To ensure understanding, full explanations are provided after submitting answers, and subjects are required to fix any incorrect responses before proceeding. At the end of the experiment, subjects are given a questionnaire on demographic information and a self-reported measure of risk-taking. Figure \ref{fig:flow} shows a flowchart of the experimental procedure.

Having Workers and Managers play identical Parts 1 and 2 serves two important purposes. First, it enables an easy interpretation of the behavior and ability signals for participants in the Manager sessions, since Managers see the same instructions and experience the same environment that produced those signals. This differs from Bartling et al. (2012), where they are careful not to reveal the details of the contracts that generated the observed behavior. In their setting, obfuscation mirrors the reality that Managers may not be aware of the contract that induced a Worker's history. To focus on how behavior and ability influence Worker evaluations, I instead opt for complete information about the signal-generating process so that a Manager need not conjecture on the conditions under which the observed signals were generated. Second, these measures allow for inference as to how the trade-off between Worker characteristics is related to the Manager's own traits. Since this design eliminates biases in meeting opportunities by having all Managers evaluate the same set of Workers, any extra value placed on a Worker's trait is likely driven by biases in preferences\footnote{Currarini et al. (2009) study homophily in friendship formation, but note that biases in meeting opportunities and preferences influence homophily in social networks more broadly.} (Currarini et al., 2009). Including the Manager's outcomes from Parts 1 and 2 as control variables in the statistical analysis provides information on the extent that Managers place extra value on dimensions they themselves are stronger in, acting as a test for homophily. 

A total of 20 subjects participated in the Worker session and 96 subjects participated across 5 Manager sessions. No subjects could participate in both session types. Participants were predominantly a mix of undergraduate and graduate students. Sessions lasted approximately 1 hour, with average earnings around \$21 CAD. All sessions were held at the McMaster Decision Science Laboratory (McDSL) at McMaster University in Hamilton, Canada. This experiment was preregistered on OSF\footnote{The preregistration can be found at https://osf.io/n89q5.}.

\subsection{Design}

\subsubsection{Part 1: Measuring Prosocial Behavior}

Subjects first play a modified Dictator Game. All subjects are randomly paired off, with one randomly assigned the role of Dictator and the other player acting as the Receiver. The Dictator is endowed with 10 tokens to divide; the Receiver has no action. Before knowing their role, all participants decide how many tokens they would send to their counterpart, as if they were the Dictator (strategy method; Fehr and Fischbacher, 2004). Roles are then announced and the Dictator's allocation is realized. Using this simple strategy method enables the collection of token allocation choices for Receivers without concern for history or order effects induced by having each subject play in each role.

\begin{figure}[h]
    \centering
    \includegraphics[width=0.333333\linewidth]{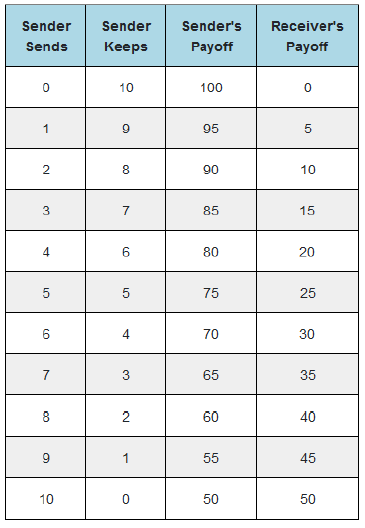}
    \caption{Screenshot of Dictator Game Table}
    \label{fig:DGTable}
\end{figure}

Each token the Dictator sends reduces their payoff by 5 points and increases the Receiver's payoff by 5 points, up to a 50/50 allocation. Specifically, tokens retained earn the Dictator 10 points each, while tokens shared earn both the Dictator and the Receiver 5 points, and sending all tokens amounts to an equal share of points. This payoff structure resembles a public goods game with two players a single contributor, where retaining tokens is analogous to investing in the private asset and sending tokens is analogous to contributing. Figure \ref{fig:DGTable} shows a screenshot of the payoff table participants observed. 

An even split appears to be a behavioral upper bound in Dictator Game sharing, with practically no offers above 50\% of the available surplus (Falk and Fischbacher, 2006). This design essentially replaces the strategically irrelevant choices with relevant ones so that the full action space is used, which is helpful in creating variance in send amounts. Although this modified structure could have an impact on the distribution of sent tokens relative to a standard Dictator Game, the purpose of this task is to generate any suitable distribution of token choices. A participant's decision in a Dictator Game encapsulates concerns for fairness, inequality aversion, altruism, or other prosocial considerations and is thus treated as a general signal of such instrinsic motivations.

Pairs who complete Part 1 before others wait to proceed to Part 2 so that no information about Part 2 is leaked to those still engaged in Part 1. Let $x_n \in \{0, \dots, 10\}$ be the amount of tokens sent by participant $n$. 

\subsubsection{Part 2: Measuring Ability}

Once all subjects complete Part 1, they proceed to an independent math task, which provides a relevant signal of ability for the simulated jobs. Importantly, the math task is played after the Dictator Game to eliminate any strategic sharing considerations, such as offsetting poor performance in the task by retaining more tokens later if the Dictator Game was played second. 

\begin{figure}[h]
    \centering
    \includegraphics[width=0.5\linewidth]{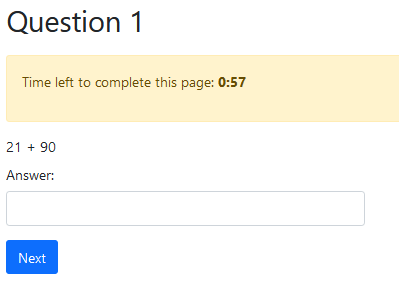}
    \caption{Screenshot of Addition Task}
    \label{fig:mathtask}
\end{figure}

Participants answer 10 two-digit addition problems with a timer continuously counting down from 60 seconds. Subjects earn 10 points for each correct answer and zero for incorrect answers. Unanswered questions are considered incorrect. Figure \ref{fig:mathtask} shows a screenshot of an example problem subjects faced. Let $y_n \in \{0, \dots, 10\}$ be the number of correct answers by participant $n$. 

\subsubsection{Part 3: Jobs Game}

In part 3, the main part of the experiment, Workers perform two jobs, and Managers evaluate Workers for these jobs given the Worker's allocation decision in Part 1 and success rate in Part 2. The two jobs, called Conflict (C) and No Conflict (NC), combine Parts 1 and 2 of the experiment in an intuitive manner. For each job, Workers decide how many addition problems to attempt out of 10, skipping the rest of them. Problems in both C and NC are roughly the same difficulty as those in Part 2. Workers are given 6 seconds times the number of questions they choose to answer, keeping the time pressure proportional to Part 2. For example, if a Worker chose to answer 5 problems for a job in Part 3, they would get 30 seconds to do so. Table \ref{tab:jobreturns} shows the parameters of the jobs.

\begin{table}[h]
\sffamily
\centering
\resizebox{.75\textwidth}{!}{%
\begin{tabular}{cccccc}
\hline
\rowcolor[HTML]{ADD8E6} 
\multicolumn{6}{|c|}{\cellcolor[HTML]{ADD8E6}\textbf{Conflict}} \\ \hline
\rowcolor[HTML]{ADD8E6} 
\multicolumn{1}{|c|}{\cellcolor[HTML]{ADD8E6}} & \multicolumn{1}{c|}{\cellcolor[HTML]{ADD8E6}Manager} & \multicolumn{1}{c|}{\cellcolor[HTML]{ADD8E6}Worker} & \multicolumn{1}{c|}{\cellcolor[HTML]{ADD8E6}} & \multicolumn{1}{c|}{\cellcolor[HTML]{ADD8E6}Manager} & \multicolumn{1}{c|}{\cellcolor[HTML]{ADD8E6}Worker} \\ \hline
\multicolumn{1}{|c|}{\begin{tabular}[c]{@{}c@{}}Answer \\ Correctly\end{tabular}} & \multicolumn{1}{c|}{10} & \multicolumn{1}{c|}{10} & \multicolumn{1}{c|}{Skip} & \multicolumn{1}{c|}{0} & \multicolumn{1}{c|}{15} \\ \hline
\multicolumn{1}{l}{} & \multicolumn{1}{l}{} & \multicolumn{1}{l}{} & \multicolumn{1}{l}{} & \multicolumn{1}{l}{} & \multicolumn{1}{l}{} \\ \hline
\rowcolor[HTML]{ADD8E6} 
\multicolumn{6}{|c|}{\cellcolor[HTML]{ADD8E6}\textbf{No Conflict}} \\ \hline
\rowcolor[HTML]{ADD8E6} 
\multicolumn{1}{|c|}{\cellcolor[HTML]{ADD8E6}} & \multicolumn{1}{c|}{\cellcolor[HTML]{ADD8E6}Manager} & \multicolumn{1}{c|}{\cellcolor[HTML]{ADD8E6}Worker} & \multicolumn{1}{c|}{\cellcolor[HTML]{ADD8E6}} & \multicolumn{1}{c|}{\cellcolor[HTML]{ADD8E6}Manager} & \multicolumn{1}{c|}{\cellcolor[HTML]{ADD8E6}Worker} \\ \hline
\multicolumn{1}{|c|}{\begin{tabular}[c]{@{}c@{}}Answer \\ Correctly\end{tabular}} & \multicolumn{1}{c|}{10} & \multicolumn{1}{c|}{10} & \multicolumn{1}{c|}{Skip} & \multicolumn{1}{c|}{0} & \multicolumn{1}{c|}{0} \\ \hline
\end{tabular}%
}
\caption{Job Returns}
\label{tab:jobreturns}
\end{table}

In C, Workers earn points either by shirking or answering questions correctly, with a fixed piece rate of 10 points earned for each correct solution and 15 points earned for each skipped problem. Each correct solution also earns a Manager 10 points. Workers can only earn points for Managers by answering problems correctly; skipped problems earn a Manager zero points, and incorrect solutions earn both the Manager and Worker zero points. The only difference in NC is that there are no longer rewards to shirking, so the piece rate is sufficient to incentivize effort. Both C and NC pay the Worker the same piece rate for correct responses as in Part 2 plus earnings for Managers, mitigating the concern that Workers may use less effort when solving problems for the Manager relative to their effort in the math task which produced their ability signal. 

\noindent{\textit{Worker Session}}

Workers observe either C or NC and choose how many questions to answer and how many to skip. The two jobs are presented in a random order across Workers and completed sequentially. After choosing the number of problems to attempt, Workers solve the problems while the timer continuously counts down from the allotted time.  After completing both jobs, one is randomly selected for the Worker's payoff. These payoffs are independent from Manager evaluations and are paid to subjects at the end of the Worker session. This separation of payments enables the creation of a single database of Workers, so that all Managers evaluate the same set of Workers, and eliminates the need to recruit more subjects to participate as Workers in subsequent sessions. Workers' choices and any points earned for Managers are stored for use in the Manager sessions. 

\noindent{\textit{Manager Sessions}}

Managers complete a rank-and-value task to evaluate Workers from the Worker Session. Managers observe Parts 1 and 2 results $(x_n,y_n)$ for each of 20 Workers and use this information to evaluate them for the jobs. They receive no information about how Workers made their Part 3 decisions. Of the 20 subjects that participated in the Worker session, 6 had duplicate results, \footnote{See Table \ref{tab:worker_results} for the Worker results in Parts 1 and 2.} and 15 were ultimately used in the Manager sessions. An additional 5 pairs were generated to fill gaps and improve coverage over the possible $(x_n,y_n)$ pairs. Managers did not know which 5 these were, and were told that only human subject Workers would be used for final payoffs. 

\begin{figure}[h]
    \centering
    \includegraphics[width=0.5\linewidth]{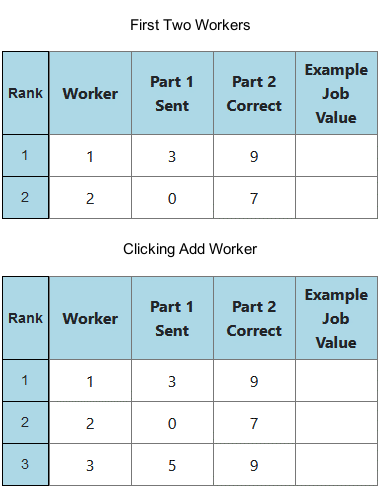}
    \caption{Table of Workers from Example Job}
    \label{fig:workertab}
\end{figure}

Evaluating Workers is done through a rank-and-value task. Managers observe Worker measures in a provided table with columns named Rank, Worker, Part 1 Sent, Part 2 Correct, and Job Value. They also observe job C or NC. Initially, only two Workers appear in the table. Subjects click and drag the row of their preferred Worker for the job to Rank 1, the top row, before clicking the ``Add Worker'' button to add the next Worker. This creates a new row in the table representing the new Worker to be ranked. This sequential approach combined with an ability to place similarly-ranked Workers near each other in the table allows Managers to compare Workers in the simplest way. Figure \ref{fig:workertab} shows an example of the table of Workers that Managers observe. Workers and jobs are presented in a randomized order across Managers to limit order effects on evaluations.

Subjects proceed with sequentially ranking the Workers until all 20 are ranked. Once the ranking is complete, Managers are instructed to enter a value from 0 to 100, the lowest and highest number of points a Worker can earn them, for each Worker in the Job Value box. They are told to think about their value as some number of points such that, if they were offered more, they would prefer the points, and if they were offered less, they would prefer their earnings to be determined by the Worker. Subjects are free to enter the same value for two Workers if they are indifferent between them. Thus, reports aim to elicit the Manager's value of each Worker in points\footnote{Equivalently, in CAD, since points are ultimately converted to CAD at the end of the experiment.}, producing a cardinal comparison of Workers.  This rank-and-value task is repeated for the remaining job once evaluations for the first job are complete. This results in a total of 40 evaluations per Manager, with 20 for each job.

To elicit accurate reports, the Manager is first randomly assigned two finalist Workers for one randomly selected job. Managers are told that the Worker with the higher reported value is their preferred Worker and will be considered for payoffs. This makes the Worker comparison salient to the Manager during the valuation process. The value of the preferred Worker is then compared to a random number $\alpha \in [0,100]$ as in the Becker-DeGroot-Marschak mechanism (Becker et al., 1964), with bounds equal to the minimum and maximum points a Worker can earn for a Manager in either job. If $\alpha$ is lower than the reported value of the Worker, the Manager earns payoffs according to the Worker's performance in the job. Otherwise, they earn $\alpha$ points, reflecting a failure to hire where $\alpha$ may be interpreted as an exogenously determined reservation utility.

This payoff structure is incentive compatible. To see this, let $V$ be the true value and $\tilde{V}$ be the reported value of the preferred finalist Worker. If $V < \alpha < \tilde{V}$, they earn payoffs according to the Worker's performance, despite their preference of receiving $\alpha$. If $V > \alpha > \tilde{V}$, the certain amount $\alpha$ is received despite the true preference of having the Worker perform the job.

\subsection{Empirical Analysis}

Values $V_n = V(x_n, y_n)$ elicited for each Worker $n$ represent a type of certainty equivalent such that the expected utility of hiring Worker $n$ is equal to the utility value of $V_n$:
\begin{align*}
    E\big[u(\pi) \lvert x_n, y_n\big] &= u\big(V_n(x_n,y_n)\big) \\
    \implies \frac{\partial E[u(\pi) \lvert x_n, y_n]}{\partial z} &= \frac{\partial u}{\partial V_n} \frac{\partial V_n}{\partial z} 
\end{align*}
where $\pi$ is a monetary payoff and $z$ is either $x$ or $y$. The analysis in this paper seeks to estimate $\frac{\partial V}{\partial x}$ and $\frac{\partial V}{\partial y}$ at various levels of $x$ and $y$ conditional on Manager characteristics. 

First, I estimate a nonparametric regression of the form 
\begin{align}
    E[V_n  \ \lvert \ x_n, y_n] = f(x_n, y_n) \label{eq:1}
\end{align}
where $f(\cdot)$ is estimated using generalized product kernels with the \texttt{np} package in \texttt{R} (Hayfield and Racine, 2008). The specification used takes advantage of the natural order of $x_n$ and $y_n$. The estimated model is used to predict reported values $\hat{V}$ at each point in the grid $G = \{ (x,y) \lvert (x,y) \in \{0,\dots, 10\} \times \{4,\dots, 10\} \}$. $G$ represents a set of representative Workers with sent amounts and math scores at each point from $(0,4)$, the minimum Worker Sent and Worker Score in the data, to $(10,10)$, the maximums. The increase in predicted value for a one-unit change in either the sent amount or the math score is then calculated as the difference in predicted values at grid points that differ only in the single dimension. For example, for the Worker at point $(x,y)$, the increase in value for a one-unit increase in the sent amount is calculated as $\hat{V}(x+1,y) - \hat{V}(x,y)$, and similarly the value increase for a one-unit increase in math scores is $\hat{V}(x,y+1) - \hat{V}(x,y)$. These local differences are visualized with heatmaps. 

To quantify the average effect globally, I estimate a linear fixed-effects model of the form:
\begin{align}
    V_{i,n} &= \alpha_i + \beta_0 + \beta_1 x_n + \beta_2 y_n + \beta_3 x_n \times y_n
    + \Phi (\text{Controls}) + \epsilon_{i,n} \label{eq:2} \\
    \epsilon_{i,n} &\sim \mathcal{N}(0, \sigma^2_\epsilon) \notag
\end{align}
where $i$ indexes Managers and $n$ indexes Workers. The fixed-effects are included to control for individual-specific intercepts in values. Controls include indicator variables for the Manager's results from Parts 1 and 2 interacted with Worker characteristics, plus some additional demographic information collected from the post-experiment questionnaire. One specification uses binary variables for whether the Manager is in the top third of token senders or math scores, while another considers whether the Manager sent more tokens than problems they solved correctly ($x_i > y_i$). Since there is no variation within Managers' own sent amounts and math scores, only interaction terms with these values can be identified. Coefficients on the interaction terms are interpreted as the extra value a Manager places on Workers' behavior or ability measures when satisfying the interacting condition. 

\subsection{Hypotheses}

\subsubsection{Worker Rankings and Valuations}

\noindent \textit{H1: Values $V$ are (weakly) higher in NC than in C.}

Jobs differ only in their payoffs to Workers for shirking. In the NC job, there are no payoffs for shirking and the Worker's interests are perfectly aligned with the Manager's under the financial incentives alone. Thus, there will be fewer skipped problems in NC than C, leading to higher overall valuations.  

\noindent \textit{H2: In the Conflict job, tokens sent $x_n$ and correct solutions $y_n$ are equally valuable to Managers, $\beta^{\text{C}}_1 = \beta^{\text{C}}_2$.}

In C, even if a Worker expects to answer all 10 questions correctly, they can earn more by simply shirking on the Manager. At the time of choosing the number of questions to attempt, the game is effectively a Dictator Game with uncertainty on the share action. To this end, individuals who have signaled past willingness to share surplus may be viewed as those willing to share surplus with the Manager by attempting questions. Even so, the Worker needs to answer the questions correctly to earn points for the Manager. Since they earn a piece-rate for doing so, it is expected the concern for generosity will be attenuated relative to the designs in Raihani and Barclay (2016) and Eisenbruch and Roney (2017). There is no obvious case ex-ante for either characteristic to be preferred. 

\noindent \textit{H3: In the No Conflict job, Workers are ranked on $y_n$ only, $\beta^{\text{NC}}_1 = 0$, or lexicographically with priority on math ability.}

Since there is no incentive to shirk in NC, Managers expect that Workers behave rationally and answer all questions. If there are still weak preferences\footnote{Specifically, if a Manager requires a sent increase of 10 units or more to offset a one unit decrease in score, a lexicographic ordering is optimal. A simple example is $U(x_n, y_n) = x_n + 11y_n$.} for $x_n$, they will rank according to $y_n$ first, and then within groups of Workers with common $y_n$, rank by $x_n$.

\subsubsection{Likeness Bias}

\noindent \textit{H4a: In both jobs, the marginal effect of $x_n$ on values is higher for high sending Managers in the Dictator Game.}

\noindent \textit{H4b: In both jobs, the marginal effect of $y_n$ on values is higher for high scoring Managers in the math task.}

Homophily in social networks has been documented in the economics and sociology literature. There is evidence to show that this phenomenon exists with job referrals (Burks et al., 2015) and that in-group biases exist (Currarini and Mengel, 2016; Robalo et al., 2017; Paetzel and Sausgruber, 2018). To the extent that birds of a feather do flock together, Managers will tend to favor Workers with characteristics similar to their own. In particular, high senders will place relatively high value on the token share amounts while Managers with high math scores will place relatively high value on math ability. As well, social-type Managers who sent more tokens than their score on the math task, $x_i > y_i$, will also place relatively more value on sent tokens and relatively less value on math scores than Managers who correctly answered at least as many math problems as the number of tokens shared.

\section{Results}

\subsection{Workers}

\begin{figure}[H]
    \centering
    \includegraphics[width=.8\linewidth]{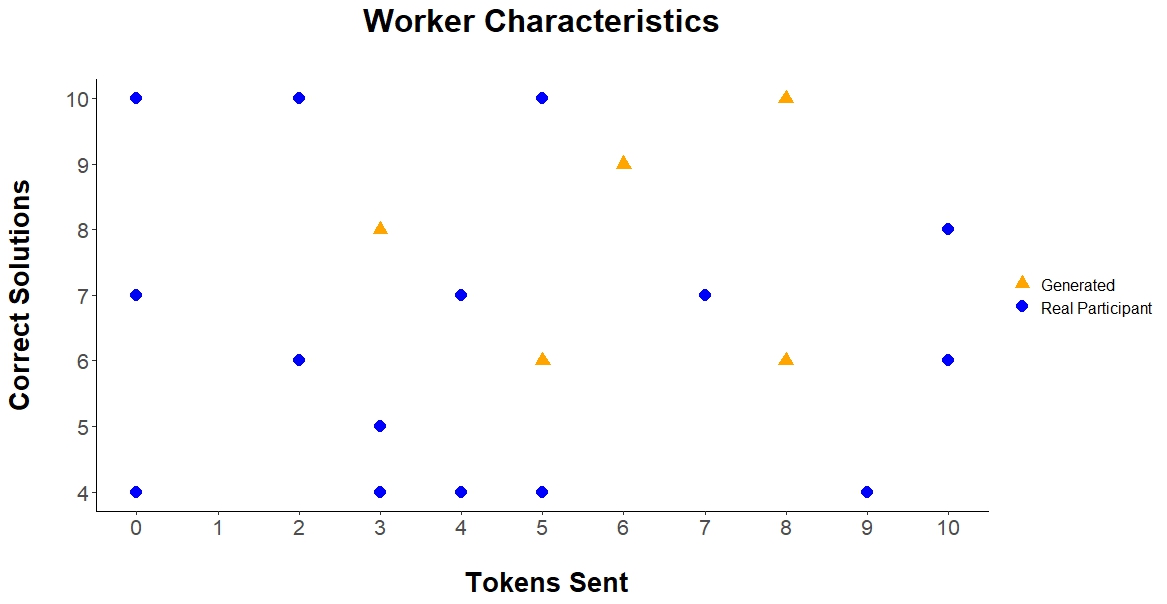}
    \caption{Workers In Pool}
    \label{fig:worker_pool}
\end{figure}

Workers were coded using IDs that were not visible to Managers and were presented in a random order. The set of evaluated Workers is given in Table \ref{tab:workerids} using the internal IDs\footnote{The original Worker results can be found in the appendix.}. Specific Workers will be referred to in this section as Worker ID $(x_n,y_n)$, explicitly stating the Worker's characteristics to ease the reading. The scatter plot in Figure \ref{fig:worker_pool} visualizes the results from Parts 1 and 2 in the Worker session and includes all Workers that Managers evaluated. From the 20 subjects in the Worker session, 15 are included in the Worker pool, represented as blue circles. The 5 excluded Workers were either duplicates or add very little to the variation in Workers (see Table \ref{tab:worker_results}). Five additional pairs were generated to create additional variance in Workers, represented by orange triangles. 

\begin{table}[h]
\centering
\resizebox{.75\textwidth}{!}{%
\begin{tabular}{|c|c|c|l|c|c|c|}
\cline{1-3} \cline{5-7}
\textbf{Worker ID} & \textbf{\begin{tabular}[c]{@{}c@{}}Tokens \\ Sent\end{tabular}} & \textbf{\begin{tabular}[c]{@{}c@{}}Correct \\ Solutions\end{tabular}} &  & \textbf{Worker ID} & \textbf{\begin{tabular}[c]{@{}c@{}}Tokens \\ Sent\end{tabular}} & \textbf{\begin{tabular}[c]{@{}c@{}}Correct \\ Solutions\end{tabular}} \\ \cline{1-3} \cline{5-7} 
1 & 3 & 5 &  & 11 & 0 & 7 \\ \cline{1-3} \cline{5-7} 
2 & 5 & 10 &  & 12 & 10 & 8 \\ \cline{1-3} \cline{5-7} 
3 & 2 & 10 &  & 13 & 0 & 10 \\ \cline{1-3} \cline{5-7} 
4 & 4 & 7 &  & 14 & 10 & 6 \\ \cline{1-3} \cline{5-7} 
5 & 9 & 4 &  & 15 & 7 & 7 \\ \cline{1-3} \cline{5-7} 
6 & 2 & 6 &  & 16 & 8 & 10 \\ \cline{1-3} \cline{5-7} 
7 & 4 & 4 &  & 17 & 5 & 6 \\ \cline{1-3} \cline{5-7} 
8 & 0 & 4 &  & 18 & 8 & 6 \\ \cline{1-3} \cline{5-7} 
9 & 5 & 4 &  & 19 & 6 & 9 \\ \cline{1-3} \cline{5-7} 
10 & 3 & 4 &  & 20 & 3 & 8 \\ \cline{1-3} \cline{5-7} 
\end{tabular}%
}
\caption{Table of Workers by ID}
\label{tab:workerids}
\end{table}

Results from the experiment are aligned with crowding-out, though the number of Workers in the sample is small (20) since the focus in this paper is on Manager evaluations. Two Workers answered all 10 questions in job C, even with insufficient financial incentives. Several Workers chose to put some effort into the jobs even when shirking paid more, with job effort increasing only slightly with no shirk temptation.  

\subsection{Managers}

A total of 96 subjects participated in the Manager sessions. From these 96 subjects, 18 are excluded from analysis, leaving 78 sets of Manager reports for study. Subjects who reported the same valuation for all Workers across the two jobs were excluded because such responses indicate a lack of attention to the differences in Worker characteristics or the job-specific returns. Additionally, a uniformly high valuation, particularly for Workers with the lowest characteristics (e.g., (0,4)), is inconsistent with the premise of the task, which assumes that Worker valuations should reflect differences in characteristics and the implied returns. Assigning near-maximum values to all Workers suggests either a misunderstanding of the task or an unwillingness to meaningfully differentiate between workers, leading to data that does not reflect genuine preferences or considerations.  Each Manager reported values for all 20 potential Workers for a total of 1,560 reports per job. 

\subsubsection{Aggregate Rankings}

\begin{figure}[H]
    \centering
    \includegraphics[width=.75\linewidth]{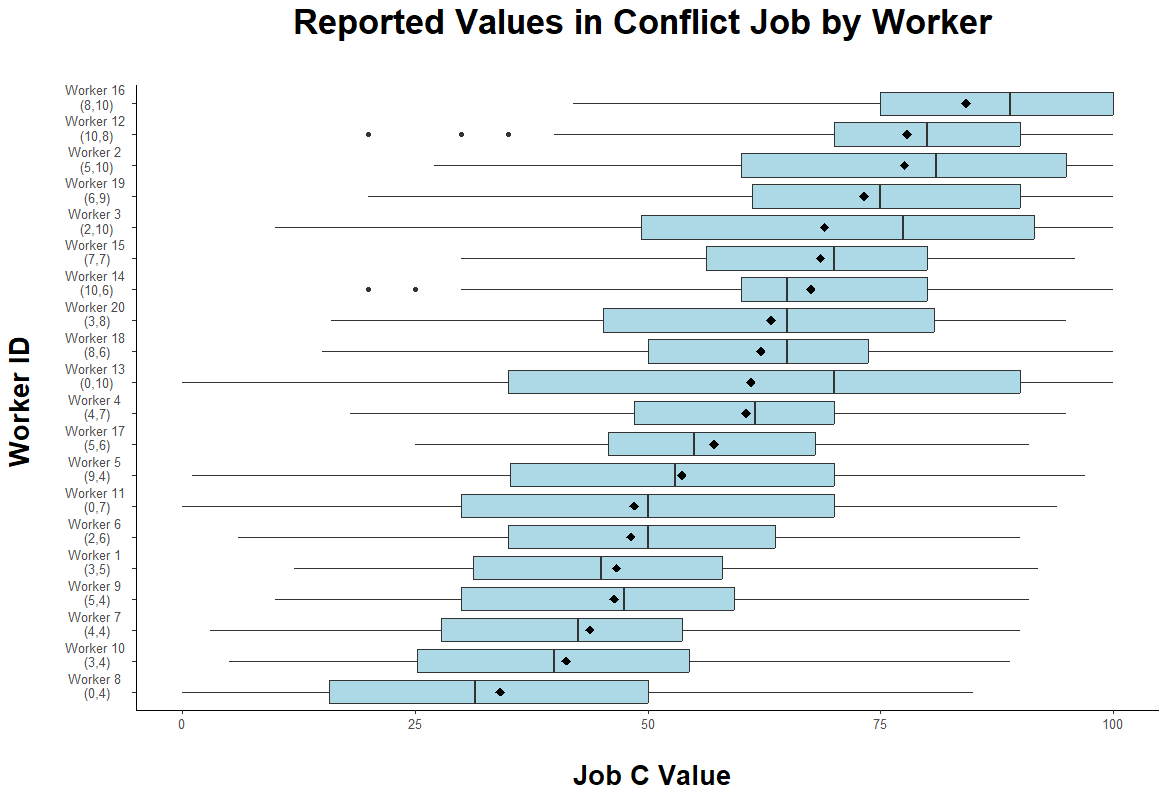}
    \caption{Box Plots of Job C Values}
    \label{fig:job1boxplot}
\end{figure}

\begin{figure}[H]
    \centering
    \includegraphics[width=.75\linewidth]{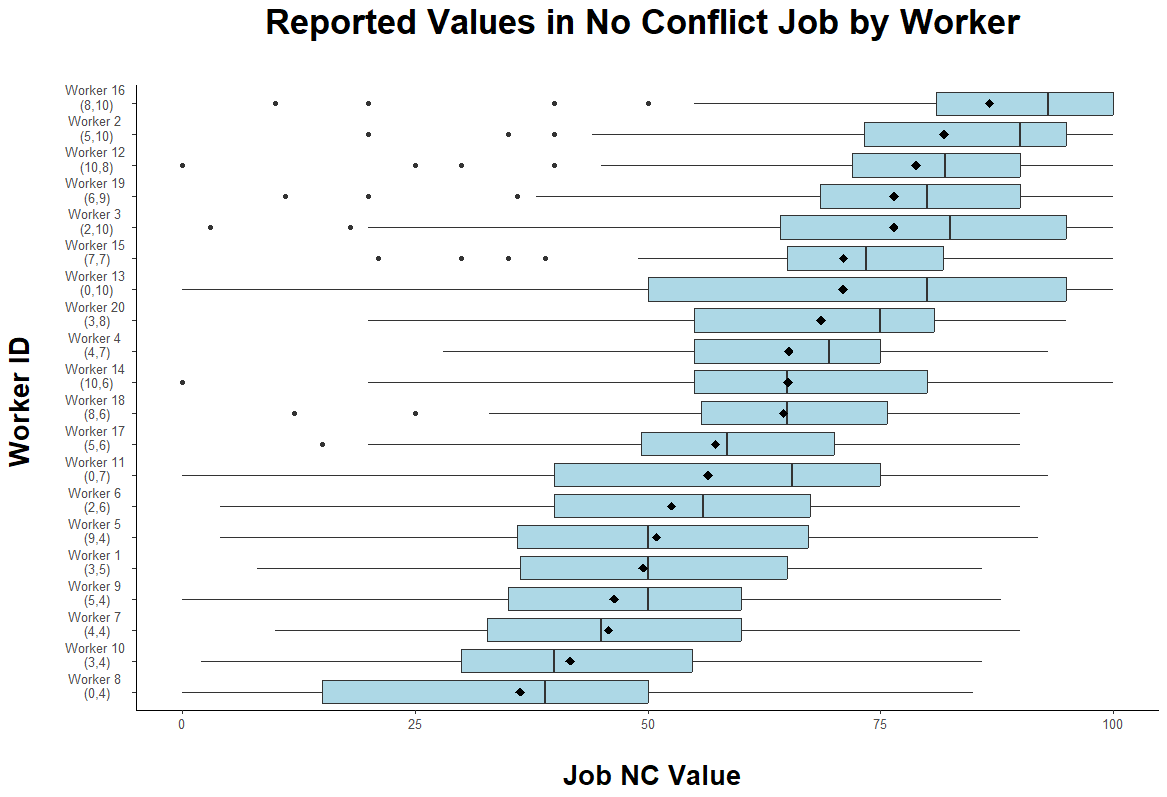}
    \caption{Box Plots of Job NC Values}
    \label{fig:job2boxplot}
\end{figure}

The valuation data reveals which Workers are generally regarded as the best and worst on average. Figures \ref{fig:job1boxplot} and \ref{fig:job2boxplot} show the boxplots of reported values for each Worker for C and NC, respectively, with mean values represented by black diamonds and Workers ranked in order of mean reported value on the y-axis. 

In C, Worker 2 $(5,10)$, Worker 12 $(10,8)$, and Worker 16 $(8,10)$ received the highest shares of values above 80 points. Worker 16 $(8,10)$, having sent 8 tokens and answered all 10 math questions correctly, had the highest average reported value with a mean of 84.2 points, significantly different than the second highest mean value of 77.9 (t-test p $= 0.023$) by Worker 12 $(10,8)$, the mirror image of Worker 16 $(8,10)$. Half of the Managers valued Worker 16 $(8,10)$ at over 90 points (39 out of 78 subjects). Worker 2 $(5,10)$, who sent less tokens but answered all of the math questions in the math task correctly, had a mean value of 77.6, not statistically different from Worker 12 $(10,8)$. Taken together, these top 3 Workers foreshadow Managers' preference toward math ability: Worker 16 $(8,10)$ is preferred on average to Worker 12 $(10,8)$, but there is no significant difference between Worker 2 $(5,10)$ and Worker 12 $(10,8)$ in aggregate.  

In NC, the top 3 Workers remain the same. The mean reported value for Worker 2 $(5,10)$ increases and becomes greater than that of Worker 12 $(10,8)$, but insignificantly so (81.9 vs 78.9, t-test p = $0.32$). This is somewhat surprising since Worker 12 $(10,8)$ answered fewer questions correctly than some workers not in the top 3. Worker 11 $(0,7)$ and Worker 13 $(0,10)$, who each sent zero tokens but performed well on the math task, increase in value significantly, by 8 points (t-test p $= 0.047$) and 10 points (t-test p $= 0.042$) respectively. Mean values for Worker 3 $(2,10)$, Worker 4 $(4,7)$, and Worker 20 $(3,8)$ increase marginally (t-tests p $\in (0.05, 0.10)$). Each of these Workers sent relatively few tokens but answered at least 7 questions correctly in the math task. Workers whose values changed the least between jobs are all in the bottom left quadrant of Figure \ref{fig:worker_pool}. Average values for Worker 12 $(10,8)$ and Worker 14 $(10,6)$, each of whom sent all 10 tokens, were also stable between jobs. Worker 8 $(0,4)$ had the lowest reported mean value for both jobs. These stability results are intuitive: low quality workers receive low average values in either job without much change, while workers that evenly share the surplus, such as Worker 12 $(10,8)$ and  Worker 14 $(10,6)$, may be expected to avoid the temptations of the gains from shirking and perform similarly across jobs as well. 

The boxplots also show disagreement in the valuations of some Workers, demonstrated by wider boxes. This is especially pronounced for Workers high in one dimension and low in the other. Worker 13 $(0,10)$ stands out in both jobs, with an interquartile range of 55 points in C and 45 points in NC. Worker 3 $(2,10)$ and Worker 11 $(0,7)$ are the only other workers with a range of 40 points or more in C. In fact, most of the disagreement is on the five Workers on the boundaries of the top-left and bottom-right quadrants of Figure \ref{fig:worker_pool}. It is these Workers with high levels in one dimension but not the other that creates the tension central to the hiring dilemma. Workers with high (low) levels in both characteristics consistently rank at the top (bottom), while these Workers with differing strengths elicit mixed responses. This pattern is aligned with the social perception literature, where individuals perceived as being both warm and competent elicit uniformly positive feelings, individuals perceived as lacking both warmth and competence elicit uniformly negative feelings, and individuals perceived as high in one dimension but low in the other elicit mixed feelings (Fiske et al., 2007). There is less disagreement in NC, though relatively flat distributions persist.

\begin{figure}[H]
    \centering
    \includegraphics[width=\linewidth]{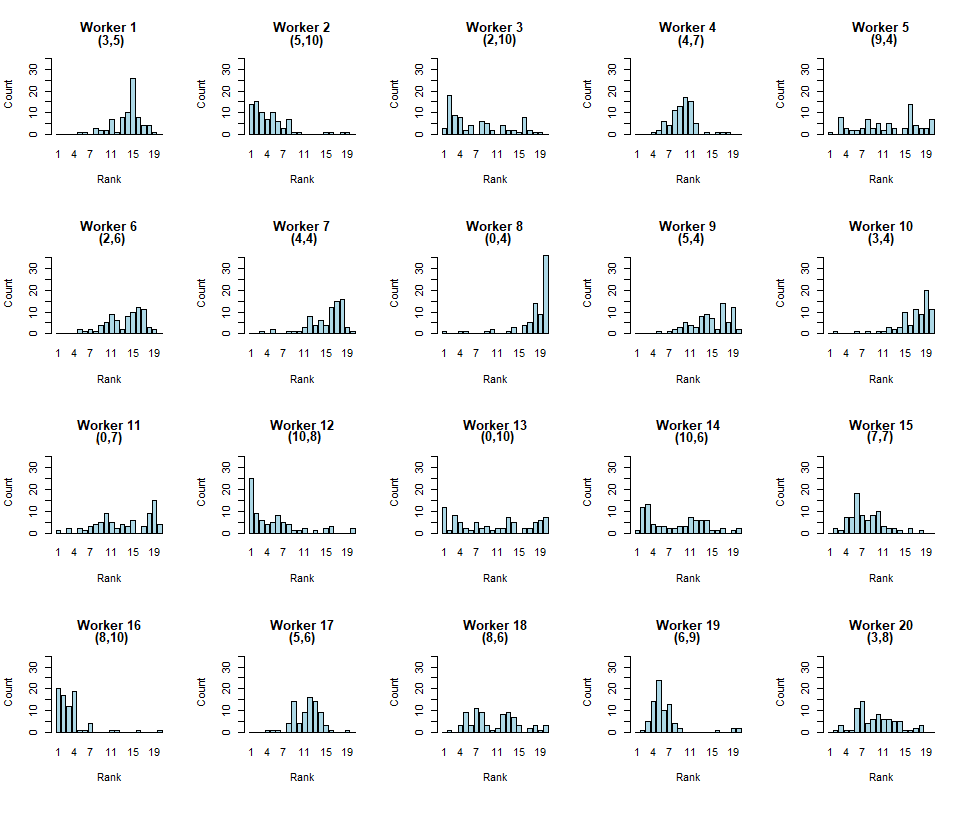}
    \caption{Rank Distribution of Workers in C}
    \label{fig:rankdist1}
\end{figure}

\begin{figure}[H]
    \centering
    \includegraphics[width=\linewidth]{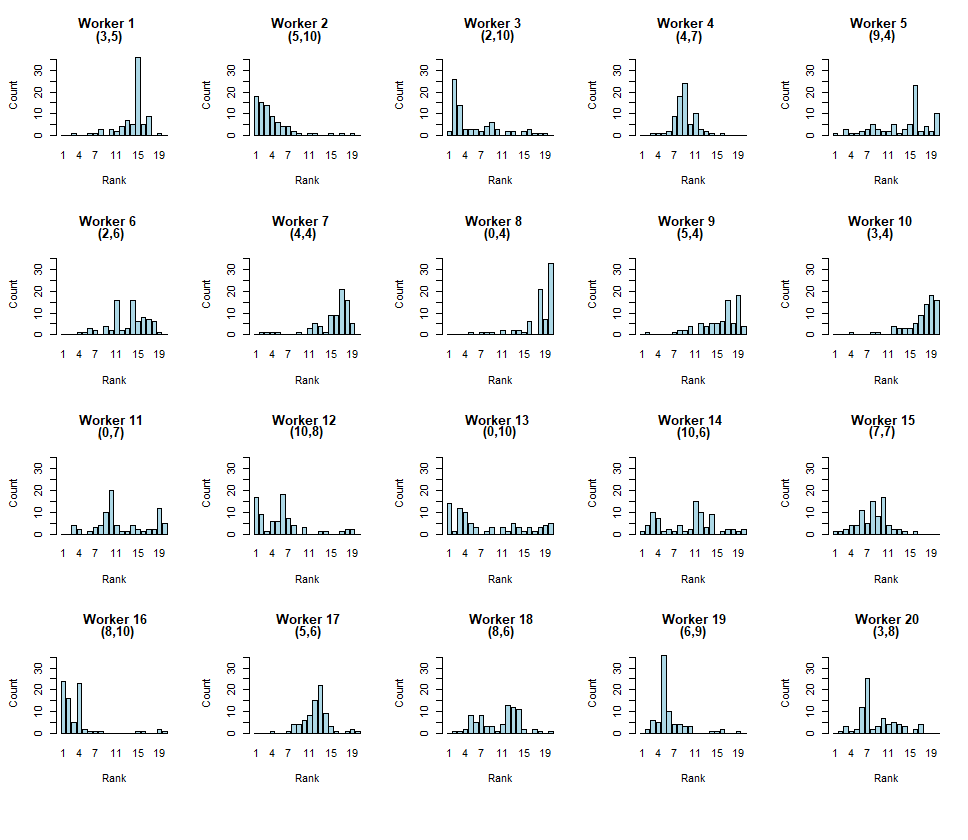}
    \caption{Rank Distribution of Workers in NC}
    \label{fig:rankdist2}
\end{figure}

Ordinal rankings implied by Managers' reports allow for rank comparison across Workers and are provided as well. Figures \ref{fig:rankdist1} and \ref{fig:rankdist2} shows the distributions of aggregate rankings for each job. The x-axis of each plot represents the rank position, with the count of Managers ranking them in that position on the y-axis. Mass toward the left side indicates high frequencies of top ranks, whereas mass on the right side indicates higher frequencies of low ranks. Looking at rankings controls for individual specific intercepts in Managers' value scales, which appears to be a significant factor given the dispersion of values observed in the boxplots.

As expected, Worker 16 $(8,10)$ consistently ranks among the top Workers. Worker 12 $(10,8)$, the mirror image of Worker 16 $(8,10)$, was ranked number 1 more often, but had more evenly distributed mass across the top 10. A similar though flatter pattern is observed for Worker 2 $(5,10)$. Worker 8 $(0,4)$ was ranked consistently at the bottom, with some ranks near the bottom simply as a result of ties. 

Flatter distributions suggest stronger disagreement on the value of the Worker in question, whereas peaks suggest agreement. For instance, Worker 19 $(6,9)$ is rarely a top 3 worker, but almost always among the top six, suggesting agreement on their overall ranking of about 5th for either job. On the other hand, 21 Managers ranked Worker 13 $(0,10)$ in the top 3 for C, with 12 of these Managers putting them in the top rank, but 22 Managers placed them among the bottom 5. Similarly, Worker 3 $(2,10)$ was ranked in the top 3 by 30 Managers and in the bottom 5 by 12 Managers. The extreme difference in characteristics for these Workers — sharing few or no tokens but achieving a perfect score in the math task — suggests a divide in preferences: individuals who prioritize math skills seem to place high value on these low-sharing Workers, while those who emphasize social characteristics tend to rank them lower.

Aggregate evidence from the experiment supports H1. In the baseline job C, the mean value reported for Workers is 59.3, slightly lower than the mean of 62.2 in the no conflict job (t-test, p $< 0.01 $), a difference worth about \$0.23 CAD. This result is driven mostly by increased values to Workers with very low sent amounts but relatively high scores.

\subsubsection{Main Results}

\begin{figure}[H]
    \centering
    \includegraphics[width=\linewidth, height=8cm]{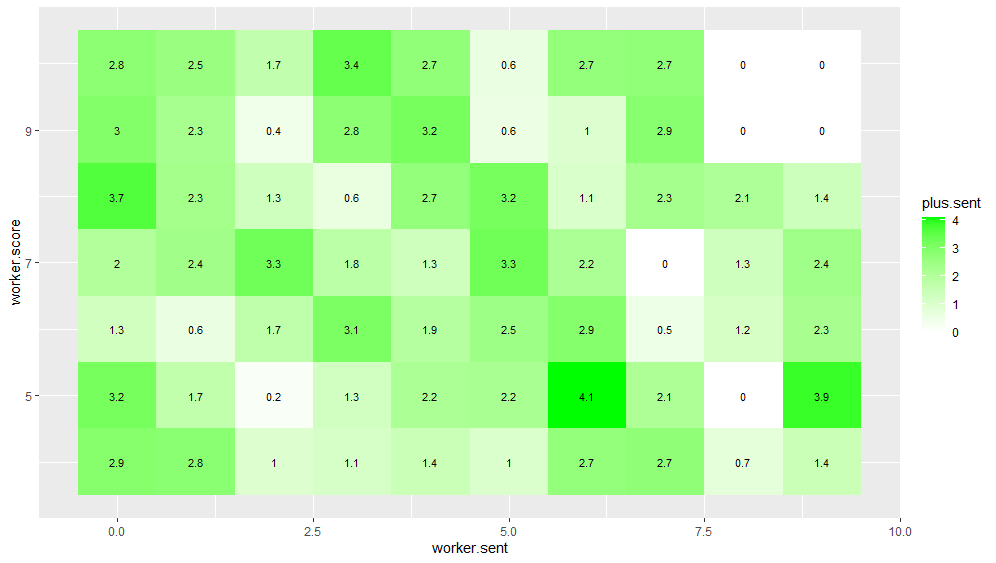}
    \caption{Increase in Reported Value for +1 Sent $V(x+1,y) - V(x,y)$}
    \label{fig:plussent}
\end{figure}

\begin{figure}[H]
    \centering
    \includegraphics[width=\linewidth, height=8cm]{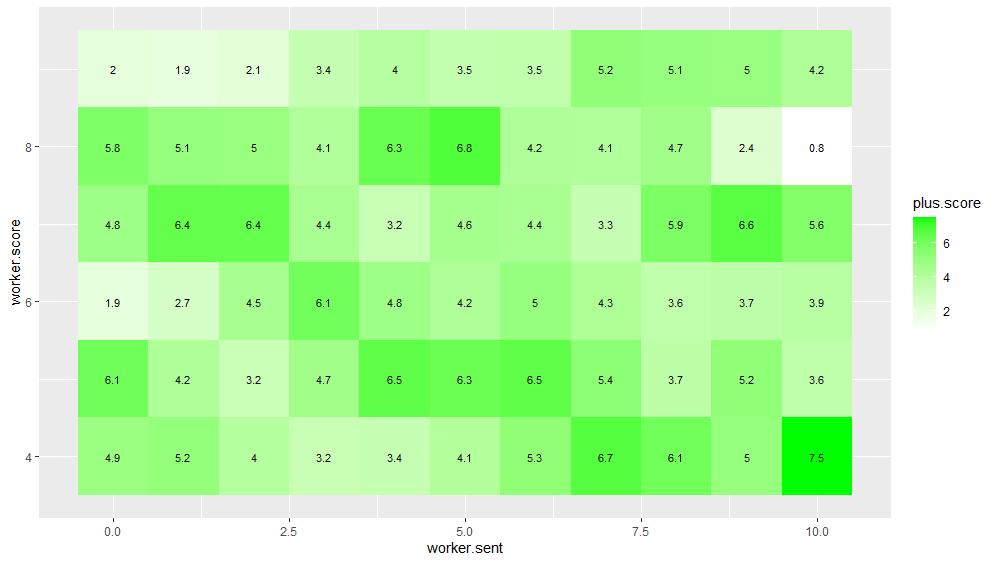}
    \caption{Increase in Reported Value for +1 Score $V(x,y+1) - V(x,y)$}
    \label{fig:plusscore}
\end{figure}

\begin{figure}[H]
    \centering
    \includegraphics[width=\linewidth, height=8cm]{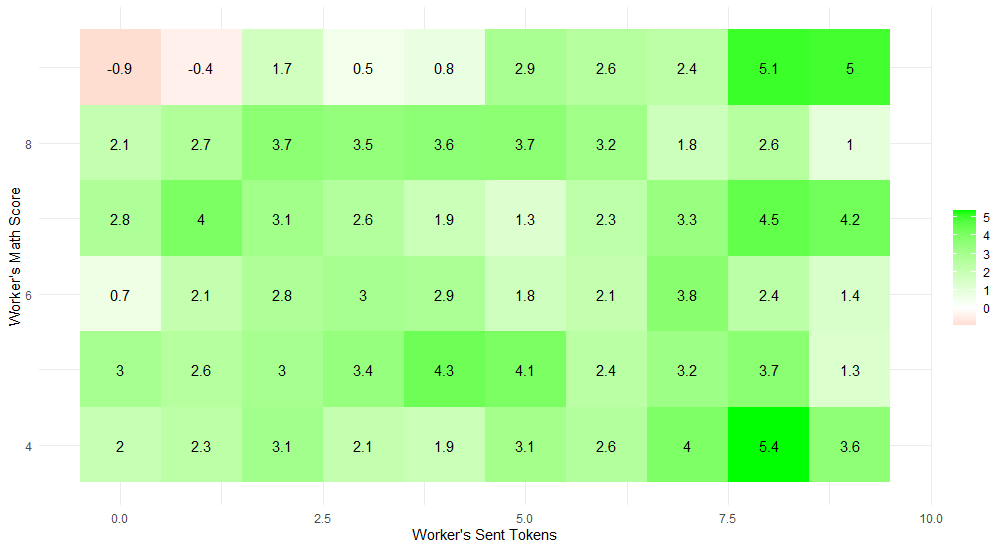}
    \caption{Difference in Increases in Reported Value $V(x,y+1) - V(x+1,y)$}
    \label{fig:plusdiff}
\end{figure}

To understand the relative value of a Worker's social or ability component to a Manager, I first estimate the relationships nonparametrically following equation (\ref{eq:1}) for job C. The results are visualized in Figures \ref{fig:plussent} and \ref{fig:plusscore}, showing the differences in increasing either $x_n$ or $y_n$ by one unit at each grid point as heatmaps. Figure \ref{fig:plusdiff} combines these heatmaps, highlighting the additional value Managers place on an extra correct solution in the math task compared to sending one more token. Values in the figure correspond to a simple double difference:
\begin{align*}
    \bigg(\hat{V}(x,y+1) - \hat{V}(x,y)\bigg) - \bigg(\hat{V}(x+1,y) - \hat{V}(x,y)\bigg) = \hat{V}(x,y+1) - \hat{V}(x+1,y) \\
\end{align*}

The results highlight a strong preference for Worker scores over Worker sent amounts, leading to rejection of H2 in general. For 58 out of 60 grid points, the increase in predicted value is substantially larger for a 1-unit increase in the Worker's math score than a 1-unit increase in sent tokens. Sending an extra token is more valuable than solving an additional problem correctly only for Workers who answered 9 questions correctly and sent 0 or 1 token. At high levels of both sent tokens and correct solutions, there is no increase in value for sending more tokens. Compared to Workers with high math scores, increases in Worker values are small for low math score Workers when increasing their sent amounts by one. Increases in correct solutions lead to relatively larger increases in value for low scoring Workers than high scoring Workers, especially low scorers who send a majority of tokens. 

To quantify the effect on average, and to investigate the extent to which this effect is influenced by the Manager's characteristics, I estimate four specifications of equation (\ref{eq:2}) for each job. Tables \ref{tab:feCreg} and \ref{tab:feNCreg} show the results. The first column of each table shows estimates from a basic specification with only the Worker's characteristics and their product. The second column adds two questionnaire controls\footnote{Other controls such as Age and self-reported risk-taking were not significant.}: an indicator variable for whether the participant is in a STEM faculty and whether they indicated their gender as Male in the questionnaire. Managers were assigned as being part of a STEM faculty if they indicated their faculty as Engineering, Sciences, or Health Sciences in the questionnaire\footnote{Faculties not in STEM were Social Sciences, Business, and Humanities.}. Specifications in columns (3) and (4) test for a likeness bias in valuations. Column (3) includes interaction terms of the Worker's characteristics with indicator variables for whether the Manager sent 6 or more tokens or answered 8 or more math problems correctly. These values correspond with the top third in each characteristic (26 and 28 out of 78 participants, respectively). Finally, column (4) includes interaction terms of the Worker's characteristics with an indicator variable for whether the Manager sent more than they scored. This implicitly categorizes Managers into two types: those with stronger prosocial motivations, and those with stronger ability.

\begin{table}[!htbp] \centering 
  \caption{Fixed-Effects Regressions for Job C} 
  \label{tab:feCreg} 
\begin{tabular}{@{\extracolsep{5pt}}lcccc} 
\\[-1.8ex]\hline 
\hline \\[-1.8ex] 
 & \multicolumn{4}{c}{\textit{Dependent variable:}} \\ 
\cline{2-5} 
\\[-1.8ex] & \multicolumn{4}{c}{Values in C} \\ 
\\[-1.8ex] & (1) & (2) & (3) & (4)\\ 
\hline \\[-1.8ex] 
 Worker Sent Amount & 1.740$^{***}$ & 2.214$^{***}$ & 1.611$^{***}$ & 1.619$^{***}$ \\ 
  & (0.435) & (0.488) & (0.505) & (0.481) \\ 
 Worker Math Score & 4.644$^{***}$ & 4.050$^{***}$ & 4.725$^{***}$ & 4.856$^{***}$ \\ 
  & (0.311) & (0.452) & (0.496) & (0.455) \\ 
 Worker Sent $\times$ Worker Score & 0.101$^{*}$ & 0.101$^{*}$ & 0.101$^{*}$ & 0.101$^{*}$ \\ 
  & (0.061) & (0.060) & (0.059) & (0.058) \\ 
 Worker Sent $\times \ \mathds{1}(\text{STEM})$ &  & $-$1.080$^{***}$ & $-$1.025$^{***}$ & $-$1.015$^{***}$ \\ 
  &  & (0.290) & (0.287) & (0.282) \\ 
 Worker Score $\times \ \mathds{1}(\text{STEM})$ &  & 0.662 & 0.588 & 0.573 \\ 
  &  & (0.418) & (0.413) & (0.406) \\ 
 Worker Sent $\times \ \mathds{1}(\text{Male})$ &  & 1.029$^{***}$ & 1.099$^{***}$ & 1.258$^{***}$ \\ 
  &  & (0.264) & (0.262) & (0.259) \\ 
 Worker Score $\times \ \mathds{1}(\text{Male})$ &  & 0.279 & 0.214 & $-$0.032 \\ 
  &  & (0.381) & (0.377) & (0.372) \\ 
 Worker Sent $\times \ \mathds{1}(\text{Manager Sent } \ge 6)$ &  &  & 1.230$^{***}$ &  \\ 
  &  &  & (0.263) &  \\ 
 Worker Score $\times \ \mathds{1}(\text{Manager Sent } \ge 6)$ &  &  & $-$1.715$^{***}$ &  \\ 
  &  &  & (0.378) &  \\ 
 Worker Sent $\times \ \mathds{1}(\text{Manager Score } \ge 8)$ &  &  & 0.357 &  \\ 
  &  &  & (0.259) &  \\ 
 Worker Score $\times \ \mathds{1}(\text{Manager Score } \ge 8)$ &  &  & $-$0.070 &  \\ 
  &  &  & (0.372) &  \\ 
 Worker Sent $\times \ \mathds{1}(\text{Manager Sent } > \text{Score})$ &  &  &  & 2.033$^{***}$ \\ 
  &  &  &  & (0.289) \\ 
 Worker Score $\times \ \mathds{1}(\text{Manager Sent } > \text{Score})$ &  &  &  & $-$2.755$^{***}$ \\ 
  &  &  &  & (0.415) \\ 
\hline \\[-1.8ex] 
Observations & 1,560 & 1,560 & 1,560 & 1,560 \\ 
R$^{2}$ & 0.631 & 0.639 & 0.649 & 0.661 \\ 
\hline 
\hline \\[-1.8ex] 
\textit{Note:}  & \multicolumn{4}{r}{$^{*}$p$<$0.1; $^{**}$p$<$0.05; $^{***}$p$<$0.01} \\ 
\end{tabular} 
\end{table}  

\begin{table}[!htbp] \centering 
  \caption{Fixed-Effects Regressions for job NC} 
  \label{tab:feNCreg} 
\begin{tabular}{@{\extracolsep{5pt}}lcccc} 
\\[-1.8ex]\hline 
\hline \\[-1.8ex] 
 & \multicolumn{4}{c}{\textit{Dependent variable:}} \\ 
\cline{2-5} 
\\[-1.8ex] & \multicolumn{4}{c}{Values in NC} \\ 
\\[-1.8ex] & (1) & (2) & (3) & (4)\\ 
\hline \\[-1.8ex] 
 Worker Sent Amount & 1.409$^{***}$ & 0.679 & 0.588 & 0.203 \\ 
  & (0.427) & (0.476) & (0.498) & (0.479) \\ 
 Worker Math Score & 5.762$^{***}$ & 6.116$^{***}$ & 6.461$^{***}$ & 6.331$^{***}$ \\ 
  & (0.306) & (0.441) & (0.490) & (0.453) \\ 
 Worker Sent $\times$ Worker Score & 0.046 & 0.046 & 0.046 & 0.046 \\ 
  & (0.060) & (0.059) & (0.059) & (0.058) \\ 
 Worker Sent $\times \ \mathds{1}(\text{STEM})$ &  & 0.159 & 0.182 & 0.212 \\ 
  &  & (0.283) & (0.283) & (0.280) \\ 
 Worker Score $\times \ \mathds{1}(\text{STEM})$ &  & $-$0.341 & $-$0.354 & $-$0.365 \\ 
  &  & (0.408) & (0.407) & (0.404) \\ 
 Worker Sent $\times \ \mathds{1}(\text{Male})$ &  & 1.827$^{***}$ & 1.823$^{***}$ & 2.010$^{***}$ \\ 
  &  & (0.258) & (0.258) & (0.257) \\ 
 Worker Score $\times \ \mathds{1}(\text{Male})$ &  & $-$0.289 & $-$0.349 & $-$0.372 \\ 
  &  & (0.371) & (0.372) & (0.370) \\ 
 Worker Sent $\times \ \mathds{1}(\text{Manager Sent } \ge 6)$ &  &  & 0.557$^{**}$ &  \\ 
  &  &  & (0.259) &  \\ 
 Worker Score $\times \ \mathds{1}(\text{Manager Sent } \ge 6)$ &  &  & $-$0.213 &  \\ 
  &  &  & (0.373) &  \\ 
 Worker Sent $\times \ \mathds{1}(\text{Manager Score } \ge 8)$ &  &  & $-$0.307 &  \\ 
  &  &  & (0.255) &  \\ 
 Worker Score $\times \ \mathds{1}(\text{Manager Score } \ge 8)$ &  &  & $-$0.683$^{*}$ &  \\ 
  &  &  & (0.367) &  \\ 
 Worker Sent $\times \ \mathds{1}(\text{Manager Sent } > \text{Score})$ &  &  &  & 1.627$^{***}$ \\ 
  &  &  &  & (0.287) \\ 
 Worker Score $\times \ \mathds{1}(\text{Manager Sent } > \text{Score})$ &  &  &  & $-$0.734$^{*}$ \\ 
  &  &  &  & (0.413) \\ 
\hline \\[-1.8ex] 
Observations & 1,560 & 1,560 & 1,560 & 1,560 \\ 
R$^{2}$ & 0.634 & 0.647 & 0.649 & 0.655 \\ 
\hline 
\hline \\[-1.8ex] 
\textit{Note:}  & \multicolumn{4}{r}{$^{*}$p$<$0.1; $^{**}$p$<$0.05; $^{***}$p$<$0.01} \\ 
\end{tabular} 
\end{table} 

Estimated coefficients $\hat{\beta}_1$ and $\hat{\beta}_2$ from equation (\ref{eq:2}) confirm the tastes for ability. In all cases in C, the magnitude is much higher for $\hat{\beta}_2$ than $\hat{\beta}_1$ (one-tailed t-test, $p < 0.01$). For three specifications, the estimate for an increase in value for a one-point increase in math score is three times the amount as for an additional sent token, even with controls. Under the richer specifications, a one unit increase in math scores translates to an increase of just under 5 points, or 40 CAD cents, whereas an extra sent token increases worker values by just over 1.5 points, or 12 CAD cents. These amounts, 40 and 12 CAD cents, represents 5\% and 1.5\% of the maximum \$8 that could be earned for each point increase in math score and token share, respectively. This suggests that, on average, Worker 13 $(0,10)$ is still worth at least half of the maximum earnings available from jobs, and worth about the same as a Worker with characteristics $(9,7)$. The additional value for increasing tokens sent or math scores is weakly related to the current level of the other characteristic in all specifications ($\hat{\beta}_3 = 0.101, \ p < 0.1$). 

When there is no incentive to shirk (job NC), Table \ref{tab:feNCreg} reveals that individuals still value sent tokens when no controls are included. The inclusion of additional regressors shows that, on average, the number of tokens a Worker sent did not impact values in a statistically meaningful way. Scores were the primary driver of values, as expected, lending support for H3. 

The likeness bias is investigated with specifications in columns (3) and (4) of each regression table. There is strong evidence for H4a, a likeness bias on the behavior component. All else equal, a Manager in the top third of token senders valued an extra token shared 76\% more than Managers who shared less (2.84 points vs 1.61). These high sharing Managers also valued score 36.5\% less than low sharing Managers (3.01 points vs 4.73). Column (4) shows that the effect is even more pronounced for Managers who sent more than they scored. For this group with higher social capital than ability, value placed on an extra sent token is 2.25 times more than those with weakly stronger math ability (3.65 points vs 1.62). This group of Managers also valued score substantially less (2.10 points vs 4.86). Taken together, prosocial behavior compensates for talent almost entirely for Managers in the top-third of senders (2.84 point increase per token shared vs 3.01 points per correct solution), and is even more valuable for Managers who sent more tokens than the number of correct solutions they submitted in the math task (3.65 point increase per token shared vs 2.10 points per correct solution). I find no evidence that high scoring individuals, or individuals who answered at least as many questions correctly as the number of tokens they sent, value a Worker's math score any higher than individuals who scored in the bottom two thirds in the math task or sent more than they scored, leading to rejection of H4b. 

This bias also explains why sent amounts still mattered to some Managers in NC. Even in this environment, higher sending individuals appear preferred to some low senders. Worker 13 $(0,10)$ is one of only 4 Workers to answer all 10 math problems correctly in Part 2, but sending 0 tokens resulted in being ranked 7th by mean value reports. Worker 12 $(10,8)$ and Worker 19 $(6,9)$, who scored 8 and 9 on the math task respectively, are still ranked higher than Worker 13 $(0,10)$ for more than half of the Managers. The high rank for these workers in this job appear to be explained by Managers who sent a relatively high number of tokens themselves. Managers in the top third of sent amounts placed a small but statistically significant premium on token shares. Interestingly, top scoring Managers tended to place \textit{less} value on an additional unit of score in NC than lower scoring Managers, though the effect is marginally significant ($p = 0.063$). Sending more tokens than the number of correct math problems is also significantly related to placing higher values on workers with higher social capital. 

In all, the likeness bias result fails to show up for score since all Managers have a strong taste for ability, regardless of their own characteristics. Managers placing value on a worker's social component are mainly those who had a high social score themselves. In C, participants in a STEM faculty valued sent amounts less than in other faculties, while non-STEM males appeared to value it more. The effect of STEM disappears in NC, though Males still tended to place excess value on sent tokens. 

\subsubsection{Additional Results - The Dictator Game}

\begin{figure}[H]
    \centering
    \includegraphics[width=\linewidth]{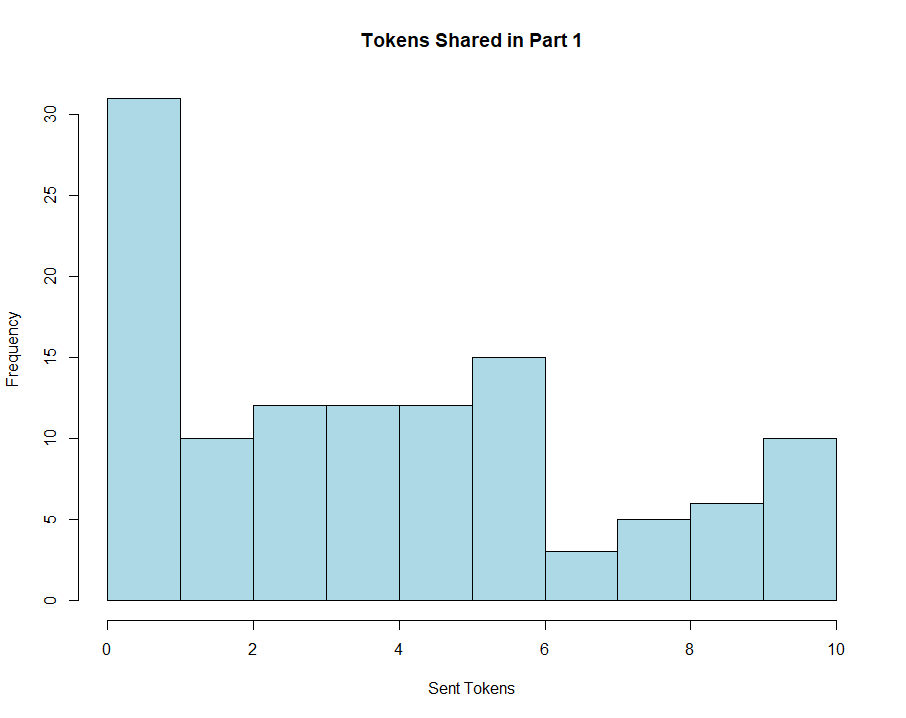}
    \caption{Histogram of Token Sent Amounts in Part 1}
    \label{fig:DGhist}
\end{figure}

Results from Dictator Games tend not to be robust to treatment manipulations (Falk and Fischbacher, 2006). In general, the most commonly observed distributions of sent tokens have two mass points: one at zero, and one at 50\% of surplus, with a distribution of sent amounts below the 50\% threshold. Results from 96 Managers and 20 Workers in the modified Dictator Game used in part 1 do not feature the commonly observed mass point at 50\%. Instead, there is a tendency to share 50\% of \textit{tokens} or less, with a relatively even distribution from 1 to 6 tokens. Frequency picks up again at the 10 token mark, representing a 50\% surplus share. Figure \ref{fig:DGhist} shows the overall distribution. 

The results suggest there is some effect due to the cutoff of surplus sharing at 50\%, despite lack of observed share amounts above this level in standard Dictator Games. The distribution of sent tokens in this study suggests that Dictators tend to favor sending zero or half of the tokens, regardless of their payoff consequences. Only a single large mass point appears at zero, with a relatively uniform distribution of offers from 5\% to 50\% of surplus except for a drop in frequency of sending 7, 8, or 9 tokens. While the distribution is important for this study and not its shape, these results contribute to the evidence of the lack of robustness of the Dictator Game, and cautions on its use without this understanding. 

\section{Discussion of Design Choices}

In practice, an Agent's history is signaled by some measure of reputation. Knowing the value of their reputation, Agents have an incentive to strategically manipulate their signals (Levashina et al., 2014; Ball, 2024). There is evidence to show that individuals use altruistic signals as a tool to improve their chances at being selected for a task (Barclay, 2004) and that when this deception is possible, others view their behavior skeptically (Barclay and Willer, 2007). Since the focus of this paper is to understand the relative value of prosocial behavior and ability signals, design choices are made to alleviate the concern that subjects may view signals as being generated strategically, so that signals can be assessed according to subjects' preferences.

The optimal mix of ability and behavioral characteristics depends on the contract and job environment itself. The job was performed in only a single period; stronger reputation concerns may come into play when a Worker performs multiple jobs for the Manager, shifting how these traits may be balanced. As well, one can imagine a situation where the likeness bias may work in reverse. For instance, a Principal who prefers a skewed outcome but feels guilty making unfair offers can strategically delegate the bargaining decision to an Agent who is less sensitive to these psychological costs (Hamman et al., 2010). The relationship between Manager and Worker characteristics is unclear without details on the job environment.

Framing the experiment as an employer-employee relationship, with external rewards contingent on behavior, lead to contrasting results relative to neutral frames, where generosity tended to matter more for partner selection than did productive ability (Raihani and Barclay, 2016; Eisenbruch and Roney, 2017). Although a Manager-Worker relationship was framed, the experiment could instead be framed as any Principal-Agent relationship with external incentives contingent on the Agent's actions. The mix of qualities an individual looks for in a coworker or a leader may differ from the mix they look for in a subordinate employee, even under the same contract and job environment. If psychological rewards are influenced not only by external incentives but also by labels, for instance, supplying effort to a ``Manager'' instead of a ``Partner'', then it seems plausible the labels themselves may influence this trade-off, independent from any contracts.

Values were elicited using the BDM mechanism, with intuitive and well-defined bounds for the randomly drawn comparison value. This enabled participants to proceed at their own pace for Part 3 and kept the experimental sessions under 60 minutes in duration for the average participant. Alternatively, other methods could be used for eliciting Principals' valuations of Workers, such as auctions or markets for these simulated Workers. An interesting experiment would be to randomly assign each Manager a Worker, and allow them to trade Workers along with transfers.

\section{Conclusion}

In this paper, I develop a novel experimental design to evaluate the trade-offs Principals face between behavior and ability when selecting Agents. This work contributes to the growing literature on productive ability versus prosocial behavior in partner choice. Findings reveal that, in general, Managers prioritize a Worker’s math ability over their history of sharing gains, especially Managers in STEM fields who value behavior relatively little, in contrast to the literature on partner choice without contracts, where generosity is often favored over competence. Relative to these studies, attaching incentives to effort appears to reduce the value of prosocial behavior, even in job C where shirking is rewarding for Workers. Contracts act as substitutes for trust, attaching incentives to behavior that may not be realized in the absence of such rewards. Contracts thus provide a degree of assurance against shirking, reducing the premium placed on internal motivations for prosocial behavior. Internal motivations also provide some assurance, but the reliance on generosity is mitigated by the presence of action-contingent rewards. 

I find evidence for a likeness bias in which high sharing individuals place relatively more value on the behavior component, contributing to the literature on homophily in social networks. Highly prosocial Managers place relatively more value on behavior, compensating for ability or even surpassing it in value. This bias can be positive, for instance providing better corporate fit with referrals (Burks et al., 2015), or have negative consequences when there are positive externalities from diversity, such as in strategic planning, delegation, or forming teams to complete heterogeneous tasks. This is especially important when Managers are unaware of their own bias, since they cannot correct for it in the screening process. If there is homophily in the evaluation process, do candidates anticipate this, and feel more confident when being assessed by an evaluator with qualities similar to themselves? Understanding how a candidate's perception of their own qualities is influenced by the qualities of their assessor is an interesting future direction for this research. 

The experimental framework developed for this study can be applied to study other types of trade-offs in multi-dimensional partner choice. A natural question is to what extent individuals are willing to work with dishonest folks, or other behaviors that are generally considered immoral. Toxic workers are costly to firms; how do individuals feel working alongside these toxic workers? Furthermore, since many hiring decisions are made by committees, understanding how groups rank candidates and how in-group members persuade each other as to who the best candidate is would provide further evidence on how a candidate's behavior is weighed relative to their ability. Finally, the framework can be extended such that the Part 3 job is a repeated game. This repeated nature more closely mimics a real job environment where a worker is expected to perform well not only once, but reasonably consistently. Reputation likely has a stronger effect in these settings, and it is not clear ex-ante whether higher ability workers who regularly shirk would be more valuable than low ability workers who consistently work.

\section{References}

\begin{singlespace}

\textbf{Alekseev, Aleksandr, Gary Charness, and Uri Gneezy}. 2017. ``Experimental Methods: When and Why Contextual Instructions are Important.'' \textit{Journal of Economic Behavior and Organization}, 134: 48-59. 

\noindent \textbf{Ball, Ian}. Forthcoming. ``Scoring Strategic Agents.'' \textit{AEJ: Microeconomics}.

\noindent \textbf{Barclay, Pat}. 2004. ``Trustworthiness and Competitive Altruism Can Also Solve The `Tragedy of the Commons'.'' \textit{Evolution and Human Behavior}, 25: 209-220.

\noindent \textbf{Barclay, Pat, and Robb Willer}. 2007. ``Partner Choice Creates Competitive Altruism in Humans.'' \textit{Proc. R. Soc. B.}, 274: 749-753.

\noindent \textbf{Battigalli, Pierpaolo and Martin Dufwenberg}. 2022. ``Belief-Dependent Motivations and Psychological Game Theory.'' \textit{Journal of Economic Literature}, 60 (3): 833-882.

\noindent \textbf{Becker, Gordon M., Morris H. Degroot, and Jacon Marschak}. 1964. ``Measuring Utility by a Single-Response Sequential Method.'' \textit{Journal of the Society for General Systems Research}, 9 (3): 226-232.

\noindent \textbf{B\'enabou, Roland, and Jean Tirole}. 2003. ``Intrinsic and Extrinsic Motivation.'' \textit{Review of Economic Studies}, 70: 489-520.

\noindent \textbf{B\'enabou, Roland, and Jean Tirole}. 2006. ``Incentives and Prosocial Behavior.'' \textit{American Economic Review}, 96 (5): 1652-1678.

\noindent \textbf{Burks, Stephen V., Bo Cowgill, Mitchell Hoffman, and Michael Housman}. 2015. ``The Value of Hiring Through Employee Referrals.'' \textit{Quarterly Journal of Economics}, 130 (2): 805-840.

\noindent \textbf{Currarini, Sergio, Matthew O. Jackson, and Paolo Pin}. 2009. ``An Economic Model of Friendship: Homophily, Minorities, and Segregation.'' \textit{Econometrica}, 77 (4): 1003-1045.

\noindent \textbf{Currarini, Sergio, Jesse Matheson, and Fernando Vega-Redondo}. 2016. ``A Simple Model of Homophily in Social Networks.'' \textit{European Economic Review}, 90: 18-39.

\noindent \textbf{Currarini, Sergio, and Friederike Mengel}. 2016. ``Identity, Homophily, and In-Group Bias.'' \textit{European Economic Review}, 90: 40-55.

\noindent \textbf{Deci, Edward L}. ``The Effects of Externally Mediated Rewards on Intrinsic Motvation.'' \textit{Journal of Personality and Social Psychology}, 18 (1): 105-115.

\noindent \textbf{Eisenbruch, Adar B., and James R. Roney}. 2017. ``The Skillful and the Stingy: Partner Choice Decisions and Fairness Intuitions Suggest Human Adaptation for a Biological Market of Cooperators.'' \textit{Evolutionary Psychological Science}, 3: 364-378.

\noindent \textbf{Falk, Armin, and Michael Kosfeld}. 2006. ``The Hidden Costs of Control.'' \textit{American Economic Review}, 96 (5): 1611–1630. 

\noindent \textbf{Falk, Armin, and Urs Fischbacher}. 2006. ``A Theory of Reciprocity.'' \textit{Games and Economic Behavior}, 54 (2): 293–315.

\noindent \textbf{Fehr, Ernst, and Urs Fischbacher}. 2004. ``Third-party Punishment and Social Norms.'' \textit{Evolution and Human Behavior}, 25: 63-87.

\noindent \textbf{Fiske, Susan T., Amy J.C. Cuddy, and Peter Glick}. 2007. ``Universal Dimension of Social Cognition: Warmth and Competence.'' \textit{Trends in Cognitive Science}, 11 (2): 77-83.

\noindent \textbf{Frey, Bruno S}. 1997. ``On the Relationship Between Intrinsic and Extrinsic Work Motivation.'' \textit{International Journal of Industrial Organization}, 15: 427-439.

\noindent \textbf{Frey, Bruno S., and Felix Oberholzer-Gee}. 1997. ``The Cost of Price Incentives: An Empirical Analysis of Motivation Crowding-Out.'' \textit{American Economic Review}, 87 (4): 746-755.

\noindent \textbf{Hamman, John R., George Loewenstein, and Roberto A. Weber}. 2010. ``Self-Interest Through Delegation: An Additional Rationale for the Principal-Agent Relationship.'' \textit{American Economic Review}, 100 (4): 1826-1846.

\noindent \textbf{Houseman, Michael, and Dylan Minor}. 2015. ``Toxic Workers.'' \textit{Harvard Business School Strategy Unit Working Paper No. 16-057}.

\noindent \textbf{Kreps, David M}. 1997. ``Intrinsic Motivation and Extrinsic Incentives.'' \textit{American Economic Review}, 87 (2): 359-364.

\noindent \textbf{Levashina, Julia, Christopher J. Hartwell, Frederick P. Morgeson, and Michael A. Campion}. 2014. ``The Structured Employment Interview: Narrative and Quantitative Review of the Research Literature.'' \textit{Personnel Psychology}, 67: 241-293.

\noindent \textbf{Lumineau, Fabrice}. 2017. ``How Contracts Influence Trust and Distrust.'' \textit{Journal of Management}, 43 (5): 1553-1577. 

\noindent \textbf{Macfarlan, Shane J., and Henry F. Lyle}. 2015. ``Multiple Reputation Domains and Cooperative Behavior in Two Latin American Communities.'' \textit{Phil. Trans. R. Soc. B.}, 370: 20150009.

\noindent \textbf{Mcpherson, Miller, Lynn Smith-Lovin, and James M. Cook}. 2001. ``Birds of a Feather: Homophily in Social Networks.'' \textit{Annual Review of Sociology}, 27: 415-444.

\noindent \textbf{Paetzel, Fabian, and Rupert Sausgruber}. 2018. ``Cognitive Ability and In-Group Bias: An Experimental Study.'' \textit{Journal of Public Economics}, 167: 280-292.

\noindent \textbf{Raihani, Nichola J., and Pat Barclay}. 2016. ``Exploring the Trade-off Between Quality and Fairness in Human Partner Choice.'' \textit{Royal Society Open Science}, 3: 160510.

\noindent \textbf{Robalo, Pedro, Arthur Schram, and Joep Sonnemans}. 2017. ``Other-Regarding Preferences, In-Group Bias and Political Participation: An Experiment.'' \textit{Journal of Economic Psychology}, 62: 130-154.

\noindent \textbf{Weber, Libby L., and Kyle J. Mayer}. 2011. ``Designing Effective Contracts: Exploring the Influence of Framing and Expectations.'' \textit{Academy of Management Review}, 36 (1): 53-75.
\end{singlespace}
\newpage
\section{Appendix}

\subsection{}
\begin{table}[h]
\centering
\caption{Parts 1 and 2 Results from Worker Session}
\label{tab:worker_results}
\resizebox{.75\textwidth}{!}{
\begin{threeparttable}
\begin{tabular}{|c|c|c|c|c|c|c|}
\cline{1-3} \cline{5-7}
\textbf{Worker} & \textbf{Tokens Sent} & \textbf{Correct Solutions} &  & \textbf{Worker} & \textbf{Tokens Sent} & \textbf{Correct Solutions} \\ \cline{1-3} \cline{5-7} 
1 & 0 & 4 &  & 11 & 3 & 5 \\ \cline{1-3} \cline{5-7} 
2 & \cellcolor[HTML]{FFCCC9}0 & \cellcolor[HTML]{FFCCC9}4 &  & 12 & 4 & 4 \\ \cline{1-3} \cline{5-7} 
3 & \cellcolor[HTML]{FFCCC9}0 & \cellcolor[HTML]{FFCCC9}5 &  & 13 & 4 & 7 \\ \cline{1-3} \cline{5-7} 
4 & \cellcolor[HTML]{FFCCC9}0 & \cellcolor[HTML]{FFCCC9}5 &  & 14 & 5 & 4 \\ \cline{1-3} \cline{5-7} 
5 & \cellcolor[HTML]{FFCCC9}0 & \cellcolor[HTML]{FFCCC9}6 &  & 15 & 5 & 10 \\ \cline{1-3} \cline{5-7} 
6 & 0 & 7 &  & 16 & \cellcolor[HTML]{FFCCC9}5 & \cellcolor[HTML]{FFCCC9}10 \\ \cline{1-3} \cline{5-7} 
7 & 0 & 10 &  & 17 & 7 & 7 \\ \cline{1-3} \cline{5-7} 
8 & 2 & 6 &  & 18 & 9 & 4 \\ \cline{1-3} \cline{5-7} 
9 & 2 & 10 &  & 19 & 10 & 6 \\ \cline{1-3} \cline{5-7} 
10 & 3 & 4 &  & 20 & 10 & 8 \\ \cline{1-3} \cline{5-7} 
\end{tabular}
\begin{tablenotes}
\tiny
\item Shaded areas represent Workers not used in the Manager sessions database.
\end{tablenotes}
\end{threeparttable}
}
\end{table}

\end{document}